\documentclass[sigconf, nonacm]{acmart}

\usepackage{graphicx}
\usepackage{amsthm}
\usepackage{amsmath}
\usepackage{dsfont}
\usepackage{hyperref}
\usepackage{svg}

\newcommand\vldbyear{2026}
\newcommand\vldbworkshop{QC\&DKM 2026 - 2nd Workshop on Quantum Computing and Data/Knowledge Management}
\newcommand\vldbauthors{\authors}
\newcommand\vldbtitle{\shorttitle} 
\newcommand\vldbavailabilityurl{}
\newcommand\vldbpagestyle{plain}

\newtheorem{theorem}{Theorem}
\newtheorem{lemma}{Lemma}
\theoremstyle{definition}
\newtheorem{definition}{Definition}
\newtheorem{example}{Example}

\newcommand{\clF}{\mathrm{cl}_{\mathcal{F}}}
\newcommand{\rmx}{\mathrm{rmax}}
\newcommand{\vr}{\mathrm{var}}
\newcommand{\supp}{\mathrm{supp}}
\newcommand{\Tr}{\mathrm{Tr}}

\begin{document}
\title{Quantum Information-Theoretical Size Bounds for Conjunctive Queries with Functional Dependencies}

\author{Valter Uotila}
\affiliation{%
  \institution{University of Helsinki}
  \country{Finland}
}

\author{Jiaheng Lu}
\affiliation{%
  \institution{University of Helsinki}
  \country{Finland}
}

\begin{abstract}
Deriving formulations to estimate worst-case size bounds for conjunctive queries under various constraints has been at the core of theoretical database research. If the problem has no constraints or has a single functional dependency, tight worst-case size bounds are computable. If the problem has more than one functional dependency, computing tight bounds can be difficult in practice and may even require an infinite number of linear inequalities in its optimization formulation. While these challenges have been addressed with varying methods, no prior research has employed quantum information theory to address this problem. In this work, we establish a connection between earlier classical information theory-based works and quantum information theory. We propose replacing the classical Shannon entropy formulation with the quantum Rényi entropy of order $\alpha \in (0,1)$ whose entropy cone is characterized simply by non-negativity. The first key result is to express the bound in terms of optimizing over quantum states and Rényi entropy. Optimizing with respect to quantum states rather than classical distributions transfers the hardness into the problem of characterizing classical states, yielding a sound but generally not tight upper bound. We further quantify this hardness explicitly by proposing a dichotomy theorem: if the query satisfies a head absorption rule, then the Rényi program value is at most one and $|Q(D)| \leq \mathrm{rmax}(D)$. Otherwise, the program is unbounded.
\end{abstract}

\maketitle

\pagestyle{\vldbpagestyle}
\begingroup\small\noindent\raggedright\textbf{VLDB Workshop Reference Format:}\\
\vldbauthors. \vldbtitle. VLDB \vldbyear\ Workshop: \vldbworkshop.\\ 
\endgroup
\begingroup
\renewcommand\thefootnote{}\footnote{\noindent
This work is licensed under the Creative Commons BY-NC-ND 4.0 International License. Visit \url{https://creativecommons.org/licenses/by-nc-nd/4.0/} to view a copy of this license. For any use beyond those covered by this license, obtain permission by emailing \href{mailto:info@vldb.org}{info@vldb.org}. Copyright is held by the owner/author(s). Publication rights licensed to the VLDB Endowment. \\
\raggedright Proceedings of the VLDB Endowment. 
ISSN 2150-8097. \\
}\addtocounter{footnote}{-1}\endgroup

\ifdefempty{\vldbavailabilityurl}{}{
\vspace{.3cm}
\begingroup\small\noindent\raggedright\textbf{VLDB Workshop Artifact Availability:}\\
The source code, data, and/or other artifacts have been made available at \url{\vldbavailabilityurl}.
\endgroup
}

\section{Introduction}

Recent years have witnessed the application of quantum computing to databases, which is an emerging interdisciplinary field that holds potential for enhancing data management. Quantum algorithms have been applied to various database problems, such as query optimization 
\cite{DBLP:conf/vldb/KesarwaniH24,Kittelmann2024Card,Uotila2024QueryMetrics}, index tuning \cite{Gruenwald2023Index,Barbosa2024QRLIT,Trummer2024Index,Kesarwani2024Index}, and transaction scheduling \cite{Groppe2021TSGrover,Bittner2020IDEAS,Bittner2020OJCC}. In this paper, we investigate how worst-case size bounds for conjunctive queries can be characterized through concepts from quantum information theory, what insights can be gained from such formulations and what kind of limits we face. We focus on conjunctive queries, the natural class corresponding to SQL's SELECT–FROM–WHERE queries with conjunctive predicates. Computing the worst-case bound is useful for optimizing query processing and resource management in relational database systems \cite{Zhang_Mayer_Khamis_Olteanu_Suciu_2025}.

One of the most significant contributions to this problem is the AGM bound, introduced by Atserias, Grohe, and Marx \cite{doi:10.1137/110859440}. The AGM bound provides a tight theoretical upper limit on the size of the output of a join query as a function of the sizes of the relations in the query body. In particular, they leverage the fractional edge cover of the query's hypergraph to derive a bound that is optimal for queries without additional constraints. However, the AGM bound does not apply to conjunctive queries with functional dependencies (FDs), as FDs introduce additional constraints that alter the query's output size. In the presence of FDs, alternative approaches are needed. For example, Gottlob, Lee, Valiant, and Valiant (GLVV)~\cite{10.1145/2220357.2220363} proposed an approach that uses Shannon entropy to derive a linear program for computing the bound. However, the complete set of linear constraints to compute a tight bound may be infinite, and characterizing the exact set of constraints remains an open problem. 

In this paper, we introduce a quantum Rényi entropy-based approach to address the problem of estimating worst-case size bounds. We show that the Rényi entropy satisfies estimates similar to those previously developed with Shannon entropy in the context of conjunctive queries. We then define what functional dependencies mean in the context of quantum states. The intuition behind various concepts in this work is presented in \autoref{table:example}. We demonstrate how to compute an upper bound for the worst-case scenario using Rényi entropy. We then explore the limits of this formulation by presenting a dichotomy theorem that classifies the behavior of this optimization problem.

\begin{table*}[t]
\centering
\begin{tabular}{lll}
\toprule
\textbf{Database} & \textbf{Classical information theory} & \textbf{Quantum information theory} \\
\midrule
tuple $t$ over $\vr(Q)$
  & element of the support of $X$
  & tuple basis vector $|t\rangle$ (\autoref{def:tuples_in_density_matrices}) \\
result table $Q'(D)$
  & distribution $p$ on the tuples
  & classical state $\rho = \sum_t p(t)\,|t\rangle\langle t|$ (\autoref{def:tuples_in_density_matrices}) \\
body relation, projection $\pi_{u_j}$
  & marginal distribution $X_{u_j}$
  & reduced state $\rho_{u_j} = \Tr_{\vr(Q)\setminus u_j}(\rho)$ \\
worst-case instance
  & uniform distribution, $H_1 = \log|Q(D)|$
  & maximally mixed state, $H_\alpha = \log|Q(D)|$ \\
functional dependency $A \to B$
  & $H_1(X_B \mid X_A) = 0$ (\autoref{lemma:random_variable_functional_dependency})
  & $H_\alpha(\rho_{A \cup B}) = H_\alpha(\rho_A)$ for classical $\rho$ (\autoref{lemma:functional_dependency_with_states}) \\
---
  & every distribution is classical
  & constraint on classicality $H_\alpha(\Delta(\rho)) = H_\alpha(\rho)$ (\autoref{thm:certificate}) \\
worst-case exponent $s(Q)$ in \eqref{eq:sQ}
  & Shannon program (\autoref{thm:glvv_over_whole_cone})
  & R\'enyi program in \autoref{thm:main_result} with value $h_\alpha(Q)$ \\
\bottomrule
\end{tabular}
\caption{Correspondence between the database, classical, and quantum concepts in this
work.}
\label{table:example}
\end{table*}

The main contributions of this paper are summarized as follows:

\begin{itemize}
    \item We review the classical theories of deriving worst-case size bounds for conjunctive queries and present concrete examples that demonstrate differences between the previous formulations. Based on this, we develop a formal connection between databases, conjunctive queries, and quantum information theory.
    \item We present a theoretical formulation for obtaining worst-case size bounds for conjunctive queries with quantum Rényi entropy stated in~\autoref{thm:main_result}. The bound is sound but not necessarily tight, which we also show.
    \item We present dichotomy~\autoref{thm:dichotomy} that characterizes the difference between the proposed formulation and its tightness. This result also helps us understand how Rényi entropy performs on this particular task compared to previous Shannon-entropy formulations.
\end{itemize}

The paper is organized as follows. First, we review the minimum necessary prerequisites on conjunctive queries, classical and quantum information theory, and entropy cones. Then, we present the main results from the previous research. After that, we develop the connection between the database and quantum information theories, building on previous works, and prove the two main results. Finally, we discuss the results and conclude the paper. The computational examples in this paper are available at \cite{valter2024sizebounds}.

\textbf{Related research.} Previous studies on quantum computing for databases have addressed multiple research questions. One of the first problems was multiple query optimization \cite{Trummer2016MQQ} on quantum annealers, which was later addressed with gate-based devices \cite{Fankhauser2023MQQ}. Join order optimization has been the most studied problem \cite{Winker2023QMLJOO,Winker2023Tut,Calikyilmaz2023Opp,Franz2024HypeQCE,Nayak2023BiDEDE,Nayak2024DBSpektrum,schoenberger:23:qdsm,Saxena2024JOO,Schoenberger:2023:sigmod,schoenberger:23:pvldb,uotila2025leftdeepjoinorderselection}, where the proposed methods have included quadratic unconstrained binary optimization and quantum reinforcement learning. Other common topics have been cardinality estimations \cite{Kittelmann2024Card,Uotila2024QueryMetrics}, index tuning \cite{Gruenwald2023Index,Barbosa2024QRLIT,Trummer2024Index,Kesarwani2024Index}, storage design \cite{Littau24Storage}, transaction scheduling \cite{Groppe2021TSGrover,Bittner2020IDEAS,Bittner2020OJCC}, task allocation \cite{10148143}, relational deep learning \cite{Vogrin2024RelationalDL} and schema matching \cite{Fritsch2023Schema}. 

\section{Prerequisites}\label{sec:prerequisites}

This section briefly presents the necessary background on conjunctive queries, classical Shannon information theory, quantum computing, quantum information theory, and both classical and quantum entropy cones. 

\subsection{Conjunctive queries}\label{sec:conjunctive_queries}

For database theory, we use the notation from \cite{10.1145/2220357.2220363}. A database is a tuple $D = (\mathcal{U}_D, R_1, \ldots, R_m)$ which consists of a finite universe $\mathcal{U}_D$ and relations $R_1, \ldots, R_m$ over $\mathcal{U}_D$. A conjunctive query is defined as 
\begin{displaymath}
    Q := R_0(u_0) \leftarrow R_{i_1}(u_1) \wedge \ldots \wedge R_{i_n}(u_n),
\end{displaymath}
where $u_j$ is a variable list of length $|u_j|$. The variable lists are not necessarily distinct from each other. Every variable that appears in the query head $u_0$ has to appear in the body of the query. The set of all variables in the query is $\mathrm{var}(Q) = \left\{X_1, \ldots, X_n \right\}$. The single relation $R_i$ might occur multiple times in the same query $Q$. Additionally, we define that $\mathrm{rmax}(D)$ is the largest relation among $R_1, \ldots, R_m$.

Evaluating the query $Q$ on a database instance $D$ means we obtain a structure $(\mathcal{U}_D, R_0)$ so that the relation $R_0$ contains those tuples that satisfy the following criteria. The relation $R_0$ consists of all tuples $\theta(u_0) := (\theta(X) \colon X \in u_0)$, where $\theta$ ranges over all functions $\theta \colon \vr(Q) \to \mathcal{U}_D$ satisfying $\theta(u_j) \in R_{i_j}(D)$ for every $j = 1, \ldots, m$. Using this $\theta$, we define that evaluating the query $Q$ on a database instance $D$ produces the result relation
\begin{displaymath}
    Q(D) := \big\{ \theta(u_0) \mid \theta \colon \vr(Q) \to \mathcal{U}_D,\ \ \theta(u_j) \in R_{i_j}(D), \ \forall j \in [m] \big\}.
\end{displaymath}
We denote that $|Q(D)|$ is the number of tuples in the resulting relation, i.e., the cardinality.

We also write $Q' := R_0'(\vr(Q)) \leftarrow R_{i_1}(u_1) \wedge \ldots \wedge R_{i_m}(u_m)$ for the query obtained from $Q$ by replacing the head $u_0$ with the full variable list $\vr(Q)$, so that $Q'(D)$ is defined by the same displayed formula with $u_0$ replaced by $\vr(Q)$. Since $u_0 \subseteq \vr(Q)$, projecting each tuple of $Q'(D)$ onto $u_0$ recovers $Q(D)$.


This work focuses on conjunctive queries on databases with arbitrary functional dependencies. We assume the reader is familiar with the attributes of relations that identify columns. If $V$ is a list of attributes on $R$ and $t \in R$ is a tuple, then $t[V]$ contains the values in $V$-positions of $t$. If $A$ and $B$ are attributes or lists of attributes of relation $R$, a functional dependency $A \to B$ on relation $R$ means that for each pair of tuples $t_1, t_2 \in R$ holds that if $t_1[A] = t_2[A]$, then $t_1[B] = t_2[B]$. 

Following the notation in \cite{10.1145/2220357.2220363}, we also define a functional dependency between query variables. If $A$ and $B$ are variables in a query appearing in positions $i$ and $j$ respectively, and $R[i] \to R[j]$ is a functional dependency, we write that $A \to B$ is a functional dependency on $R$. 

Finally, we define a functional dependency over random variables in the same way as in \cite{DBLP:journals/corr/GogaczT15}. Let $X$ be a discrete random variable obtaining values in the set $\mathcal{X}$ and distributed by the probability distribution $p \colon \mathcal{X} \to [0, 1]$. If $X$ is a random variable whose support is a set of rows with attributes in a database $D$, then $X_{Y}$ is the restriction of $X$ to the attributes $Y$. In other words, $X_Y$ is the random variable whose support is the set of rows with attributes in $Y$, and the image $\mathrm{Im}(X_{Y})$ is the table assigning a probability for each row with attributes in $Y$. This way, we can define a functional dependency on random variables, which is a key building block between databases and classical information theory.

\begin{definition}[Functional dependency on random variables]\label{def:functional_dependency_on_random_variables}
A random variable $X$ satisfies a functional dependency $A \to B$ if the table $\mathrm{Im}(X)$ satisfies the functional dependency $A \to B$.
\end{definition}

\subsection{Quantum computing}
The definitions in this subsection are based on \cite{Nielsen_Chuang_2010}, which is the standard introduction to quantum computing. We assume that the reader is familiar with the very basics of quantum computing, such as qubits and quantum logical gates.

\begin{definition}[Density operator]\label{def:density_matrix}
For each $i = 1, \ldots, n$, assume that a quantum system can be in the states $| \varphi_i \rangle$ (not necessarily orthogonal), with corresponding probabilities $p_i$. The density operator for the system is given by
\begin{displaymath}
    \rho = \sum_{i = 1}^{n}p_i |\varphi_i \rangle \langle \varphi_i |,
\end{displaymath}
where $p_i \geq 0$ for all $i$ and $\sum_{i=1}^n p_i = 1$. Another name for a density operator is a density matrix. 
\end{definition}

We call a particular subset of density operators classical: density matrix $\rho$ is classical if it is diagonal in a tensor product basis of an $n$-party system \cite{Linden_Mosonyi_Winter_2013}. Similarly, any classical distribution can be expressed as a diagonal density matrix on a tensor product basis. We call the tensor product basis the computational basis. We want to emphasize that there are density operators that are not diagonal in a tensor-product basis, so the classical states form a proper subset of all states.



\begin{definition}[Partial trace and reduced density operator]\label{def:partial_trace}
Let $\mathcal{H} = \mathcal{H}_1 \otimes \cdots \otimes \mathcal{H}_n$ be a tensor product of Hilbert spaces, indexed by $[n] := \{1, \ldots, n\}$. For $U \subseteq [n]$ write $\overline{U} := [n] \setminus U$. The partial trace over $\overline{U}$ is defined as
\begin{displaymath}
    \Tr_{\overline{U}}\big(|a\rangle\langle b|\big)
    = \left( \prod\nolimits_{i \in \overline{U}} \langle b_i | a_i \rangle \right) |a_U\rangle\langle b_U| .
\end{displaymath}
For a density operator $\rho$ on $\mathcal{H}$ and $U \subseteq [n]$, the \emph{reduced density operator} on the subsystems indexed by $U$ is $\rho_U := \Tr_{\overline{U}}(\rho)$. In addition, we say that the subsystems indexed by $\overline{U}$ have been traced out. For two parties, $A$ and $B$, this definition recovers the standard bipartite case $\rho_{A} = \Tr_{B}(\rho_{AB})$.
\end{definition}


To identify classical states, we presented the following definition, which is based on \cite{PhysRevLett.116.120404,preskill_qcqi_lecture_notes,ibm_quantum_channel_basics}.

\begin{definition}[Pinching map]\label{def:pinching_map}
We define a linear mapping between density operators, called a pinching map (or also completely dephasing channel), as $\Delta \colon \mathcal{L}(\mathcal{H}) \to \mathcal{L}(\mathcal{H})$ such that
\begin{displaymath}
    \Delta(\rho) = \sum_{i=1}^{n} \langle i | \rho| i \rangle | i \rangle \langle i |,
\end{displaymath}
where the sum is formulated so that $| i \rangle$ forms the computational basis.
\end{definition}

\subsection{Shannon and Rényi entropies}
Many classical information-theoretical concepts generalize to quantum information theory. In this subsection, we review the background of Shannon and Rényi entropies. Shannon information theory \cite{6773024} is necessary to understand one of the main results regarding polymatroid and entropic worst-case size bounds for conjunctive queries \cite{10.1145/2220357.2220363}. Because Shannon entropy can be viewed as a classical special case of Rényi entropy $H_{\alpha}$, when $\alpha = 1$, we denote Shannon entropy as $H_{1}$.

\begin{definition}[Shannon entropy]\label{def:Shannon_entropy}
Let $X$ be a discrete random variable obtaining values in the set $\mathcal{X}$ and distributed by the probability distribution $p \colon \mathcal{X} \to [0, 1]$. The Shannon entropy is defined
\begin{displaymath}
    H_1(X) = - \sum_{x \in \mathcal{X}}p(x)\log p(x).
\end{displaymath}
\end{definition}

Conditional Shannon entropy is needed to encode functional dependencies.

\begin{definition}[Conditional Shannon entropy and chain rule]\label{def:conditional_shannon_entropy}
Let $X$ and $Y$ be discrete random variables obtaining values in sets $\mathcal{X}$ and $\mathcal{Y}$. The conditional Shannon entropy is defined as
\begin{equation*}
    H_1(X \mid Y) = -\sum_{x \in \mathcal{X}}\sum_{y\in \mathcal{Y}}p(x,y)\log(p(x \mid y)).
\end{equation*}
The conditional Shannon entropy satisfies the so-called chain rule
\begin{displaymath}
    H_1(X \mid Y) = H_1(X, Y) - H_1(Y).
\end{displaymath}
\end{definition}



Rényi entropy \cite{renyi1961measures} generalizes these entropies even further.
\begin{definition}[Rényi entropy]\label{def:renyi_entropy}
For a density operator $\rho$ and $\alpha \in (0, \infty) \setminus \left\{ 1 \right\}$, the Rényi entropy is defined by
\begin{equation}
    H_{\alpha}(\rho) = \frac{1}{1 - \alpha} \log\mathrm{Tr}\rho^{\alpha}
\end{equation}
for a fixed but otherwise arbitrary logarithm base $b > 1$.
\end{definition}

Rényi entropy generalizes other entropies when $\alpha$ approaches $1$, $0$, and $\infty$ \cite{10.1063/1.4838856,Linden_Mosonyi_Winter_2013}. If $\alpha \to 0$, the Rényi entropy coincides with Hartley (or max-entropy): $H_{0}(\rho) = \log\mathrm{rank}(\rho)$. If $\alpha \to 1$, the Rényi entropy coincides with von Neumann entropy: $H_{1}(\rho) = - \mathrm{Tr}(\rho\log\rho)$. If $\alpha \to \infty$, the Rényi entropy is called min-entropy $H_{\infty}(\rho) = - \log ||\rho||$. 

The differences among the Shannon, von Neumann, and Rényi entropies are apparent in the properties each satisfies. They are all non-negative and have minimum and maximum values \cite{10.1063/1.2165794}. The von Neumann and Rényi entropies are upper-bounded at the maximally mixed states. This is analogous to the Shannon entropy being upper-bounded by uniform distributions. The Shannon and von Neumann entropies satisfy subadditivity \cite{Araki_Lieb_1970}, strong subadditivity \cite{Lieb_Ruskai_1973}, and the triangle inequality \cite{Araki_Lieb_1970}, while Rényi entropy does not satisfy any of these \cite{Linden_Mosonyi_Winter_2013}.
\subsection{Entropy cones}
Entropy cones play a crucial role in this work, as they define the space in which the worst-case size bound optimization is performed. We define the entropy cones for each $\alpha \geq 0$ \cite{Linden_Mosonyi_Winter_2013}. Let $\mathcal{P}_{n}$ be the powerset of $[n] = \left\{1, \ldots, n\right\}$ excluding the empty set $\emptyset$.
\begin{definition}
The set of all (quantum) entropic vectors is
\begin{displaymath}
       \Sigma_{\alpha}^{n} = \left\{(H_{\alpha}(\rho_{U}))_{U \in \mathcal{P}_{n}} \mid \rho \text{ state} \right\} \subseteq \mathds{R}^{2^{n} - 1}.
\end{displaymath}
The entropic cone is the set's topological closure $\overline{\Sigma_{\alpha}^{n}}$.
\end{definition}

For the Shannon entropy, we obtain the cone with respect to the classical probability distributions.
\begin{definition}
The set of all Shannon entropic vectors is
    \begin{align*}
        \Gamma^{n} &= \left\{ (H_1(\rho_{U}))_{U \in \mathcal{P}_{n}} \mid \rho \text{ classical state} \right\} \\ 
        &= \left\{ (H_1(p_{U}))_{U \in \mathcal{P}_{n}} \mid p \text{ prob. dist.} \right\}.
    \end{align*}
Similarly, the Shannon entropy cone is the set's topological closure $\overline{\Gamma^{n}}$.
\end{definition}

The famous unsolved problem in Shannon information theory is the characterization of $\overline{\Gamma^{n}}$ in terms of inequalities that these vectors' components should satisfy. The inequalities are divided into Shannon inequalities and non-Shannon inequalities. It has been shown that there are infinitely many independent non-Shannon inequalities \cite{Matus_2007}. Considering the von Neumann entropy, the studies have shown that the characterization of $\overline{\Sigma_{1}^{n}}$, i.e., the entropy cone for von Neumann entropy, might be even harder \cite{Linden_Winter_2005} since there exist inequalities which depend on relations on the entropies of certain reduced states. Surprisingly, characterizing the entropy cone for Rényi entropy, when $0 < \alpha < 1$, is much easier \cite{Linden_Mosonyi_Winter_2013}.

\begin{theorem}[Rényi entropy cone for $0 < \alpha < 1$]\label{thm:renyi_entropy_cone}
Let $0 < \alpha < 1$. For every non-negative vector $v \in \mathds{R}^{2^n - 1}_{\geq 0}$ and every $\varepsilon > 0$, there exists a state $\rho$ so that $v$ can be arbitrarily well approximated by the Rényi entropies of the $2^n - 1$ reduced states of $\rho$, i.e., $|H_{\alpha}(\rho_U) - v_U| < \varepsilon$ for all $U \in \mathcal{P}_n$. This means
\begin{displaymath}
    \overline{\Sigma_{\alpha}^{n}} = \mathds{R}^{2^n - 1}_{\geq 0}.
\end{displaymath}
In other words, no non-trivial inequalities constrain the Rényi entropies for fixed $0 < \alpha < 1$.
\end{theorem}
\begin{proof}
Theorem 3 in \cite{Linden_Mosonyi_Winter_2013}.
\end{proof}
\section{Classical worst-case size bounds}\label{sec:worst_case_classical_bounds}
The previous theoretical research on deriving worst-case size bounds can be divided into two lines depending on whether the estimations use functional dependencies or other constraints. The first line is to study join queries. The second line studies conjunctive queries in the presence of constraints. Our work aligns with the second category.

\subsection{Bounds for join queries}
First, we briefly review prior work on estimating size bounds for join queries.

\subsubsection{AGM bound}

Atserias, Grohe, and Marx present one of the first provable tight size bounds, the AGM bound, for join queries without functional dependencies or other constraints \cite{doi:10.1137/110859440}. An example of this type of query is
\begin{equation}\label{eq:standard_query_example}
     Q := R(x, y, z) \leftarrow R_1(x, y) \bowtie R_2(y, z) \bowtie R_3(z, x)
\end{equation}
for attributes $x$, $y$, and $z$.

Their novel solution is based on fractional edge covers on hypergraphs \cite{10.1145/2636918}. An edge cover on a hypergraph (or an ordinary graph) is a subset of edges such that every vertex in the graph is adjacent to at least one edge in the set of covering edges. The edge cover number is the size of the smallest set of edges that covers all vertices, considering all possible coverings. Edge cover is a fundamental optimization problem on graphs, and it admits a linear programming formulation. The formulation for this particular database problem is the following. Let $\mathrm{rel}(Q)$ be the set of relations and $\mathrm{attr}(Q)$ the set of attributes in query $Q$. Let $A_{R}$ be the set of attributes for relation $R \in \mathrm{rel}(Q)$. Then, $\mathrm{attr}(Q) = \cup_{R \in \mathrm{rel}(Q)} A_{R}$, and the linear program to compute the fractional edge cover for a query $Q$ is
\begin{equation}\label{lin_prog:agm}
    \begin{array}{lll}
        \mathrm{minimize}  & \sum_{R \in \mathrm{rel}(Q)} x_{R} & \\
        \mathrm{subject \ to} & \sum_{R \text{ if } a \in A_{R}} x_{R} \geq 1, & \text{for } a \in \mathrm{attr}(Q) \\
        &x_{R} \geq 0, & \text{for } R \in \mathrm{rel}(Q).
    \end{array}
\end{equation}
The edge cover number is the solution to the linear program with integer variables, whereas the fractional edge cover number is the solution with rationals. One of the main results of \cite{doi:10.1137/110859440} is the bound
\begin{equation}\label{eq:AGM_bound}
    |Q(D)| \leq \prod_{R \in \mathrm{rel}(Q)}|R(D)|^{x_R},
\end{equation}
where $x_{R}$ for each $R \in \mathrm{rel}(Q)$ is the solution to the fractional edge cover in Program \eqref{lin_prog:agm}. Moreover, the bound is tight, and the proof is an insightful application of Shearer's lemma \cite{CHUNG198623}.


\begin{example}\label{ex:example1}
Here we construct a common example \cite{doi:10.1137/110859440, 10.1145/2220357.2220363, DBLP:journals/corr/GogaczT15}. Consider the following conjunctive query with attributes $x$, $y$, and $z$ without functional dependencies or additional constraints
\begin{equation}\label{eq:query_example}
    Q := R(x, y, z) \leftarrow R_1(x, y) \wedge R_2(y, z) \wedge R_3(z, x).
\end{equation}
\autoref{fig:fractional_edge_cover_xyz} illustrates the query graph for this query. In this case, we have a normal graph because every relation in the query contains two attributes.
\begin{figure}[t]
    \centering
    \includegraphics[width=0.5\columnwidth]{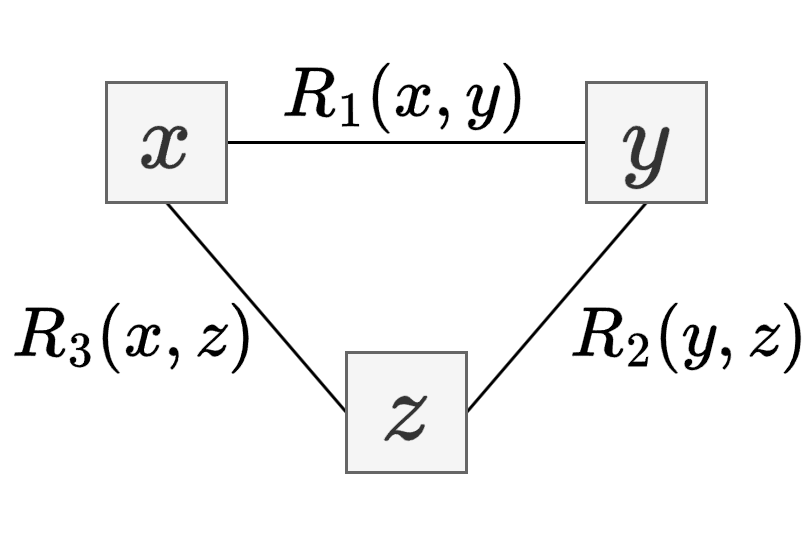}
    \caption{The example query in \autoref{eq:query_example} forms a cycle graph. The solution to the fractional edge cover assigns $1/2$ to each edge.}
    \label{fig:fractional_edge_cover_xyz}
    \Description{Query graph of three nodes representing attributes in the query. Three edges represent the relations in the query.}
\end{figure}

Using fractional edge covers on the graph, this query provides us with the following linear program to compute the AGM bound
\begin{equation*}
    \begin{array}{lll}
        \mathrm{minimize}  & x_{R_1} + x_{R_2} + x_{R_3} & \\
        \mathrm{subject \ to} & x_{R_1} + x_{R_3} \geq 1, & \text{(attribute x)} \\
         & x_{R_1} + x_{R_2} \geq 1, & \text{(attribute y)} \\
         & x_{R_2} + x_{R_3} \geq 1, & \text{(attribute z)} \\
        &x_{R_1}, x_{R_2}, x_{R_3} \geq 0, &
    \end{array}
\end{equation*}
whose solution is $x_{R_1} = x_{R_2} = x_{R_3} = 1/2$. By \autoref{eq:AGM_bound}, the tight AGM bound is
\begin{displaymath}
    |Q(D)| \leq \sqrt{|R_1|}\sqrt{|R_2|}\sqrt{|R_3|} = \sqrt{|R_1||R_2||R_3|}.
\end{displaymath}
One can see that the trivial bound is $3$, and with edge covers (no fractional), the bound is $2$. With fractional edge covers, we obtain the non-trivial worst-case size bound of $3/2$, which is also tight.
\end{example}

\subsubsection{Bounds with degree sequences}

The most recent work on estimating worst-case size bounds for join queries is \cite{khamis2025informationtheorystrikesback, Zhang_Mayer_Khamis_Olteanu_Suciu_2025, Khamis_Nakos_Olteanu_Suciu_2024, Khamis_Deeds_Olteanu_Suciu_2024}. In these works, the authors rely on previous Shannon information-theoretical size bounds and improve them by also incorporating information-theoretical inequalities on so-called degree sequences. A degree of an attribute is the number of distinct tuples in a database with that attribute. The degree sequence $d = (d_1, \ldots, d_n)$ is a sorted sequence of degrees for a collection of attributes. Various $\ell_p$-norms can be applied on these sequences defined by $||d||_p = (d_1^{p} + \cdots + d_n^{p})^{1/p}$. Depending on the used norm, one can obtain different bounds that improve the AGM bound, depending on the data in the database. While the AGM bound is tight at the relation level, optimizing the bounds at the attribute level improves it.

\subsection{Bounds for queries with functional dependencies and constraints}

\subsubsection{Polymatroid and entropic bounds}
The problem with the AGM bound is that it does not account for functional dependencies. This challenge was addressed in \cite{10.1145/2220357.2220363}, where the authors derived tight bounds for cases with no functional dependencies or with only one functional dependency. These cases were tackled with an entropy-inspired concept called a coloring number. In the general case of arbitrary functional dependencies, the authors developed a so-called polymatroid bound. They presented a conjecture that optimizing the size bound in the exact Shannon entropy cone (\autoref{def:conditional_shannon_entropy}) would provide the tight bound. This was later shown to be true \cite{DBLP:journals/corr/GogaczT15}. This tight bound is called an entropic bound. Unfortunately, the formulation for the general bound obtained in the exact Shannon entropy cone is not always practically computable in the current form because it depends on infinitely many linear inequalities in its linear program \cite{Matus_2007}.

Next, we present the background on this general case with arbitrary functional dependencies in more detail, as our quantum information-theoretical formulation is closely connected to this work. Slightly simplifying \cite{10.1145/2220357.2220363}, recall the query $Q'$ from Section~\ref{sec:prerequisites} whose head contains every variable of $Q$. The key idea is to choose a distribution $p$ on the tuples of $Q'(D)$ whose marginal on the head $u_0$ is the uniform distribution on $Q(D)$. When $Q = Q'$, as in our example \autoref{ex:example1}, every tuple is its own witness, and this reduces to the uniform distribution on all of $Q'(D)$. \autoref{ex:example2} illustrates this case. Then, the worst-case size bound can be determined by maximizing the entropy of the variables corresponding to the query head over all distributions with this property.

\begin{example}\label{ex:example2}
This example continues \autoref{ex:example1} by assigning concrete table instances for the relations in the query 
\begin{displaymath}
    Q := R(x, y, z) \leftarrow R_1(x, y) \wedge R_2(y, z) \wedge R_3(z, x).
\end{displaymath}
The tables and probabilities are visualized in \autoref{fig:example_database}. The resulting table contains the uniform distribution. A similar example was shown in \cite{10.1145/2902251.2902289}.
\begin{figure}[t]
    \centering
    \resizebox{\columnwidth}{!}{
    $
    \begin{array}{|c|c|c|c|}
    \hline
    x & y & z & X \\
    \hline
    m & 1 & u & 1/3 \\
    n & 2 & v & 1/3 \\
    m & 1 & v & 1/3 \\
    \hline
    \end{array}
    \quad
    \begin{array}{|c|c|c|}
    \hline
    x & y & X \\
    \hline
    m & 1 & 2/3 \\
    n & 2 & 1/3 \\
    \hline
    \end{array}
    \quad
    \begin{array}{|c|c|c|}
    \hline
    y & z & X \\
    \hline
    1 & u & 1/3 \\
    2 & v & 1/3 \\
    1 & v & 1/3 \\
    \hline
    \end{array}
    \quad
    \begin{array}{|c|c|c|}
    \hline
    x & z & X \\
    \hline
    m & u & 1/3 \\
    n & v & 1/3 \\
    m & v & 1/3 \\
    \hline
    \end{array}
    $
    }
    \caption{Concrete database instance continuing \autoref{ex:example1} and corresponding probabilities for the random variable $X$.}
    \label{fig:example_database}
\end{figure}
\end{example}

Since the goal is to find the worst-case size bound, we want to find $c$, such that
\begin{displaymath}
|Q(D)| \leq \mathrm{rmax}(D)^{c}.
\end{displaymath}
By solving $c$, we obtain $c \geq \log(|Q(D)|)/\log(\mathrm{rmax}(D))$. Since Shannon entropy reaches its maximum value at uniform distributions, the uniform distribution $U$ over the tuples in $Q(D)$ gives us the fact that $H_1(X) \leq H_1(U) = \log(|Q(D)|)$ for any other distribution $X$. This is the key idea why entropies are useful in these types of optimization scenarios. This leads us to aim to maximize the entropy assigned to the variables corresponding to the query head while limiting the entropy of each variable set corresponding to the relations in the query body. We will later discuss this in more detail with the Rényi entropy formulation.

Next, we discuss how to employ functional dependencies in this entropy-based formulation. The following lemma and proof are from \cite{DBLP:journals/corr/GogaczT15}. Since we prove a similar result for quantum states and Rényi entropies, we present the lemma along with its proof here. Note that the following lemma considers Shannon entropy (\autoref{def:Shannon_entropy}), denoted by $H_1$.

\begin{lemma}\label{lemma:random_variable_functional_dependency}
Random variable $X$ satisfies a functional dependency $A \to B$ if and only if $H_1(B \mid A) := H_1(X_{B} \mid X_{A}) = 0$.
\end{lemma}
\begin{proof}
Lemma~2, in~\cite{DBLP:journals/corr/GogaczT15}.
\end{proof}

The following theorem is one of the main results of \cite{10.1145/2220357.2220363} on size bounds for arbitrary functional dependencies. Let $\mathrm{rmax}(D) := \mathrm{max}_{i \in [n]}|R_i(D)|$. The following theorem expresses the size bound in terms of the query obtained by applying the chase algorithm to the original query \cite{10.1145/320083.320091}. The bound obtained from the following theorem is called a polymatroid bound.

\begin{theorem}\label{thm:glvv}
Let $Q = \textrm{chase}(Q) = R_0(u_0) \leftarrow R_{i_1}(u_1) \wedge \ldots \wedge R_{i_m}(u_m)$ be a conjunctive query with $i_j \in [m]$ and $\mathrm{var}(Q)$ be the set of variables in the query. We assume a set of arbitrary functional dependencies $\mathcal{F}$ is given. For any database $D$ satisfying $\mathcal{F}$, the following estimation holds
\begin{displaymath}
|Q(D)| \leq \mathrm{rmax}(D)^{c(Q)},
\end{displaymath}
where $c(Q)$ is the solution to the linear program in \autoref{fig:polymatroid_bound_program}. The variables of the linear program are the unconditional entropies $h(x_S)$ for all $\emptyset \neq S \subseteq [n]$.
\end{theorem}
\begin{proof}
    Proposition 6.9 in \cite{10.1145/2220357.2220363}.
\end{proof}

\begin{figure*}[t]
\begin{center}
\resizebox{0.7\textwidth}{!}{
$\begin{array}{lll}
\mathrm{maximize}  & h(u_0) & \\
\mathrm{subject \ to} & h(u_j) \leq 1 & \forall j \in [m] \\
                 & h(B \mid A) = 0 & \text{for each f.d. } A \to B \in \mathcal{F} \\
                 & h(x_j \mid x_{[n] - \left\{ j \right\}}) \geq 0 & \forall j \in [n] \\
                 & h(x_j, x_S) + h(x_l, x_S) \geq h(x_j, x_l, x_S) + h(x_S) & \forall j,l \in [n] \text{ and } S \subseteq [n] - \left\{j,l\right\}\\
\end{array}$
}
\end{center}
\caption{Linear program for \autoref{thm:glvv} to compute the polymatroid bound. The constraints implicitly assume that expressions involving mutual information or conditional entropies represent their respective linear forms of these variables.}
\label{fig:polymatroid_bound_program}
\Description{Polymatroid program represented as a figure due to long definition.}
\end{figure*}

The constraints of the linear program in \autoref{thm:glvv} and \autoref{fig:polymatroid_bound_program} can be divided into two categories. Note that $h(S)$ for any $S$ denotes the corresponding linear programming variable representing the entropy of $S$. The first category is the query-dependent constraints, which are
\begin{align}
    h(u_j) &\leq 1 \ \text{for all } j \in [m] \text{ and } \label{eq:query_dependent_constraints} \\
    h(B \mid A) &= 0 \ \text{for each f.d. } A \to B \in \mathcal{F} .\label{eq:query_dependent_constraints1}
\end{align}
We will later present the reasoning behind \autoref{eq:query_dependent_constraints}. \autoref{eq:query_dependent_constraints1} can be understood in the light of \autoref{lemma:random_variable_functional_dependency}. The second category is the information-theoretical constraints, which are
\begin{align}
    h(x_j \mid x_{[n] - \left\{ j \right\}}) &\geq 0 \text{ and } \label{eq:information_theoretical_constraints} \\
    h(x_j, x_S) + h(x_l, x_S) &\geq h(x_j, x_l, x_S) + h(x_S) \label{eq:information_theoretical_constraints2} \\
    &\forall j,l \in [n] \text{ and } S \subseteq [n] - \left\{j,l\right\} \notag.
\end{align}
The intuition behind these constraints is that the optimization should be performed with respect to variables corresponding to entropies of valid distributions. The first information-theoretical constraint is called monotonicity, and the second one is submodularity \cite{10.1145/2902251.2902289}. It is known that this formulation does not produce a tight bound if the problem involves cardinality constraints \cite{abo_khamis_what_2017}. There is also a class of inequalities, called non-Shannon inequalities \cite{681320}, which show that the constraints in \autoref{eq:information_theoretical_constraints} and \autoref{eq:information_theoretical_constraints2} are insufficient to characterize the complete entropy cone for the Shannon entropy. The question of which inequalities fully characterize this region is also an unsolved information-theoretical problem.

In \cite{DBLP:journals/corr/GogaczT15}, it was proved that if we had an exact characterization for the exact Shannon entropy cone, i.e., had a guarantee that every entropy vector in the optimization comes from a valid distribution, and utilize \autoref{eq:query_dependent_constraints} and \autoref{eq:query_dependent_constraints1}, we would obtain the tight bound. Next, we discuss this result in detail. Let
\begin{displaymath}
    s(Q, D) = \frac{\log|Q(D)|}{\log(\mathrm{rmax}(D))},
\end{displaymath}
for a database $D$ satisfying a fixed set $\mathcal{F}$ of functional dependencies. We denote
\begin{equation}\label{eq:sQ}
s(Q) = \sup_{D}s(Q, D),    
\end{equation}
where the supremum is taken over all the databases satisfying the set $\mathcal{F}$ of functional dependencies. As noted in \cite{10.1145/2220357.2220363, DBLP:journals/corr/GogaczT15}, this is the optimal value that we would like to compute in practice. The assertion in \cite{10.1145/2220357.2220363} that \autoref{thm:glvv} provides a tight bound was formally proven in \cite{DBLP:journals/corr/GogaczT15}:
\begin{theorem}\label{thm:glvv_over_whole_cone}
    Let $Q$ be a conjunctive query and $\mathrm{var}(Q)$ be the set of variables in the query. Let $\mathcal{F}$ be the set of functional dependencies. Let $H_1(Q)$ be the solution to the following linear program, when $h$ ranges over the cone $\overline{\Gamma^{n}}$:
    \begin{equation*}
    \begin{array}{lll}
    \mathrm{maximize}  & h(\mathrm{var}(Q)) & \\
    \mathrm{subject \ to} & h(\mathrm{var}(R)) \leq 1 & \text{ for } R \in \mathrm{rel}(Q) \\
                     & h(B \mid A) = 0 & \text{for each f.d. } A \to B \in \mathcal{F}.
    \end{array}
    \end{equation*}
    Then, $s(Q) = H_1(Q)$.
\end{theorem}
\begin{proof}
    Theorem 5 in \cite{DBLP:journals/corr/GogaczT15}.
\end{proof}

In this paper, we follow the naming convention in \cite{abo_khamis_what_2017} that entropic bound refers to the tight bound given in \autoref{thm:glvv_over_whole_cone}, while the polymatroid bound refers to the bound obtained with the linear program in \autoref{thm:glvv}.

Next, we still consider the connection between the Shannon entropy cone and the distributions. This also prepares the theory for considering the same construction with respect to the quantum states. The following definition was presented in \cite{DBLP:journals/corr/GogaczT15}.
\begin{definition}
Let $Q = R_0(u_0) \leftarrow R_{i_1}(u_1) \wedge \ldots \wedge R_{i_m}(u_m)$ be a query and $D$ a database satisfying a collection of functional dependencies $\mathcal{F}$. For a random variable $X$ satisfying every functional dependency $\mathcal{F}$ (in the sense of \autoref{def:functional_dependency_on_random_variables}), we define
\begin{displaymath}
    H_1(Q, X) = \frac{H_1(X_{\mathrm{var}(Q)})}{\max_{1 \leq i \leq m}H_1(X_{u_i})}.
\end{displaymath}
\end{definition}

\begin{theorem}\label{thm:optimality_with_Shannon}
The optimal result in \autoref{thm:glvv_over_whole_cone} can alternatively be obtained by taking the supremum over all the random variables $X$ satisfying the functional dependencies $\mathcal{F}$: $H_1(Q) = \sup_{X}H_1(Q, X)$.
\end{theorem}
\begin{proof}
    Proposition 14 in \cite{DBLP:journals/corr/GogaczT15}.
\end{proof}

\begin{example}
    In this example, we show how to solve \autoref{ex:example1} using the polymatroid bound in \autoref{thm:glvv}. Then, we extend the problem with functional dependencies. In both cases, the polymatroid bound is tight \cite{10.1145/2220357.2220363, 10.1145/2902251.2902289}.
    
    Even when using the polymatroid program in \autoref{fig:polymatroid_bound_program} to compute the AGM bound in the case of no functional dependencies, monotonicity and submodularity in \autoref{eq:information_theoretical_constraints} need to be included. This is clear because otherwise the variable in the objective does not depend on the variables in the constraints. Thus, the query in \autoref{ex:example1} produces the following linear program, where the variables are the unconditional entropies $h(s)$ for $\emptyset \neq s \subseteq \left\{x, y, z\right\}$:
    \begin{equation*}\label{lin_prog:polymatroid_program}
    \begin{array}{lll}
        \mathrm{maximize}  & h(xyz) & \\
        \mathrm{subject \ to} 
            & h(xy) \leq 1, & \text{(rel. xy)} \\
            & h(yz) \leq 1, & \text{(rel. yz)} \\
            & h(xz) \leq 1, & \text{(rel. zx)} \\
            & h(xyz) - h(yz) \geq 0, & \text{(monotonicity)} \\
            & h(xyz) - h(xz) \geq 0, & \text{(monotonicity)} \\
            & h(xyz) - h(xy) \geq 0, & \text{(monotonicity)} \\
            & h(xz) + h(yz) - h(xyz) - h(z) \geq 0, & \text{(submodularity)} \\
            & h(xy) + h(yz) - h(xyz) - h(y) \geq 0, & \text{(submodularity)} \\
            & h(xy) + h(xz) - h(xyz) - h(x) \geq 0, & \text{(submodularity)} \\
            & h(x) + h(y) - h(xy) \geq 0, & \text{(subadditivity)} \\
            & h(x) + h(z) - h(xz) \geq 0, & \text{(subadditivity)} \\
            & h(y) + h(z) - h(yz) \geq 0 & \text{(subadditivity)}
    \end{array}
\end{equation*}
    This linear program provides us solution $3/2$ which gives us the bound $|Q(D)| \leq \sqrt{|\mathrm{rmax}(D)|^3}$. The bound can be tighter if we have a functional dependency. For example, by including a functional dependency $\left\{ x, y \right\} \to z$, we include the additional constraint
    \begin{displaymath}
        h(z \ | \left\{ x, y \right\}) = h(xyz) - h(xy) = 0.
    \end{displaymath}
    This constraint brings the worst-case size bound to $1$.
\end{example}

Later, the polymatroid bound was further studied with lattices \cite{10.1145/2902251.2902289}, and many of the previous results were expressed in terms of lattices. In \cite{10.1145/2902251.2902289, abo_khamis_what_2017}, a new generalization for functional dependencies called degree constraints was developed. The degree constraints can be used to capture both functional dependencies and cardinality constraints. Then, it was shown that in the presence of cardinality constraints, the polymatroid bound in the linear program \autoref{fig:polymatroid_bound_program} is not tight anymore. 

Considering previous works, focusing on optimization with linear programs with modified constraints seems like a challenging task, especially since the entropy cone for Shannon entropy has been proven to be complex. Thus, we suggest that the problem should be approached with new methods. In this work, we focus on the quantum Rényi entropy but also discuss other possibilities for future research.
\section{Size bounds with Rényi entropy}

Studying whether the worst-case size bound problem can be formulated using other entropies, such as von Neumann or Rényi entropies, is a key question in this work. Unfortunately, the von Neumann entropy suffers from the same problem as the classical Shannon entropy: the exact entropy cone for the von Neumann entropy is unknown, and it is likely even more complex than the cone for Shannon entropy~\cite{Linden_Winter_2005}.

Surprisingly, the Rényi entropy for $0 < \alpha < 1$ has a simple entropy cone \cite{Linden_Mosonyi_Winter_2013}: the only restriction is non-negativity, which we stated in \autoref{thm:renyi_entropy_cone}. Any non-negative set of values assigned to the parties can be arbitrarily well approximated with $\alpha$-Rényi entropies of suitable quantum states. It is important to note that the result applies only to quantum states, and we cannot restrict ourselves to classical distributions as in \cite{10.1145/2220357.2220363}. We will later discuss the challenges that this distinction brings.

\subsection{Connecting database theory to quantum information theory}

Next, we start building a connection between the database and quantum information theories. The connection is similar to the one between databases and classical information theory, as discussed around \autoref{def:functional_dependency_on_random_variables}. Let $Q = R_0(u_0) \leftarrow R_{i_1}(u_1) \wedge \ldots \wedge R_{i_m}(u_m)$ be the given conjunctive query and $Q'$ be the query whose query head contains all the variables appearing in the query body as defined in \autoref{sec:conjunctive_queries}. We assume that the states in this section are in Hilbert space $\mathcal{H} = \mathcal{H}_{X_1} \otimes \ldots \otimes \mathcal{H}_{X_n}$ where the subspaces are indexed by the variables $\mathrm{var}(Q)$. The following definition clarifies the connection.


\begin{definition}[Encoding tuples into density matrices]\label{def:tuples_in_density_matrices}
For each variable $X_i \in \vr(Q)$, fix an orthonormal basis $\{|v\rangle \colon v \in \mathrm{dom}(X_i)\}$ of $\mathcal{H}_{X_i}$, and for a tuple $t = (v_1, \ldots, v_n)$ write $|t\rangle := |v_1\rangle \otimes \cdots \otimes |v_k\rangle$, so that $\{ |t\rangle \mid t \in Q'(D) \}$ is an orthonormal set in $\mathcal{H}$. The density matrix encoding a probability distribution $p$ over the tuples of $Q'(D)$ is then $\rho = \sum_{t \in Q'(D)} p(t)\,|t\rangle\langle t|$.
\end{definition}
The previous definition defines a mapping between the tuples and the basis of the Hilbert space. Next, we define how the partial trace can be used to model the relations that contain only part of the variables. This way, we can model the relations appearing in the query body. This is an analogous formalism to compute marginal distributions over the tuples in the classical theory.


For $S \subseteq \vr(Q)$, we write $\rho_S := \Tr_{\vr(Q) \setminus S}(\rho)$ for the reduced state on the subsystems indexed by $S$. Tracing out the variables outside $S$ is the quantum analogue of taking the marginal distribution on $S$; in particular, $\rho_{u_j}$ models the relation appearing in the $j$-th body atom.

\begin{lemma}[Matrix elements of the partial trace]\label{lem:ptrace_entries}
Let $\rho$ be a state on $\mathcal{H}_A \otimes \mathcal{H}_B$. Then, for all basis vectors $|a\rangle, |a'\rangle$ of $\mathcal{H}_A$,
\begin{displaymath}
    \langle a | \Tr_B(\rho) | a' \rangle = \sum_{b} \langle a, b | \rho | a', b \rangle .
\end{displaymath}
\end{lemma}
\begin{proof}
We can expand $\rho = \sum_{a_1,a_2,b_1,b_2} \rho_{(a_1,b_1),(a_2,b_2)}\, |a_1\rangle\langle a_2| \otimes |b_1\rangle\langle b_2|$ in the product basis. By linearity of $\Tr_B$ and its definition, we obtain $\Tr_B(|a_1\rangle\langle a_2| \otimes |b_1\rangle\langle b_2|) = |a_1\rangle\langle a_2| \,\Tr(|b_1\rangle\langle b_2|) = \delta_{b_1,b_2}\,|a_1\rangle\langle a_2|$, so $\Tr_B(\rho) = \sum_{a_1,a_2} \big(\sum_b \rho_{(a_1,b),(a_2,b)}\big) |a_1\rangle\langle a_2|$. Taking the $(a,a')$ entry gives the claim.
\end{proof}





Next, we show that with this formalism, we can model the worst-case size bounds similarly to classical theory. Whereas classically uniform distributions over the tuples represented the upper bound for the worst-case size increases, in quantum theory, the corresponding notion is the maximally mixed state. To see the connection between the Rényi-entropy and worst-case size bounds, we consider the familiar fraction
\begin{equation*}
    c = \frac{\log |Q(D)|}{\log(\mathrm{rmax}(D))}.
\end{equation*} 

First, we fix a quantum state $\rho$ over the tuples of $Q'(D)$ (containing all variables in the result) to be such that the subsystem $\rho_{u_0}$ over the values of the $|u_0|$-tuples corresponding to variables in $u_0$ is a maximally mixed state. While state $\rho$ is not necessarily unique, we assume that the reduced state has the form $\rho_{u_0} = \mathds{I}/|Q(D)|$ for an identity matrix $\mathds{I}$ of dimension $|Q(D)|$. This state exists as we constructed the witness distribution in \autoref{sec:worst_case_classical_bounds}. This reasoning is similar to that of \cite{doi:10.1137/110859440, 10.1145/2220357.2220363}. They assume a distribution $p$ such that the marginal distribution $p_{u_0}$ corresponding to the head $u_0$ is uniform, which means that for every tuple $t \in Q(D)$, the probability of selecting $t$ from the distribution $p_{u_0}$ is $1/|Q(D)|$. By \autoref{def:renyi_entropy}, we deduce that
\begin{align*}
    H_{\alpha}(\rho_{u_0}) &= \frac{1}{1 - \alpha} \log \mathrm{Tr}(\rho_{u_0}^{\alpha})\\ 
    &= \frac{1}{1 - \alpha} \log \sum_{i = 1}^{|Q(D)|} \frac{1}{|Q(D)|^{\alpha}} = \log |Q(D)|.
\end{align*}
Moreover, for any $j \in [m]$, we can see that the corresponding maximally mixed subspaces give the upper bounds
\begin{align*}
H_{\alpha}(\rho_{u_j}) &= \frac{1}{1 - \alpha}\log\mathrm{Tr}(\rho_{u_j}^{\alpha}) \\ 
&\leq  \frac{1}{1 - \alpha} \log \sum_{l = 1}^{|R_{i_j}(D)|}\frac{1}{|R_{i_j}(D)|^{\alpha}} = \log |R_{i_j}(D)|.
\end{align*}

These equalities correspond to well-known facts about Rényi entropy, which attains its maximum value at maximally mixed states, much like Shannon entropy, which reaches its maximum at a uniform distribution. Finally, combining the previous results for any $j \in [m]$, we obtain the estimate
\begin{displaymath}
\frac{H_{\alpha}(\rho_{u_0})}{H_{\alpha}(\rho_{u_j})} \geq \frac{\log(|Q(D)|)}{\log(|R_{i_j}(D)|)} \geq \frac{\log |Q(D)|}{\log(\mathrm{rmax}(D))} = c,
\end{displaymath}
which shows that we want to maximize $H_{\alpha}(\rho_{u_0})$ while limiting $H_{\alpha}(\rho_{u_j}) \leq 1$, which will bound the exponent $c$. Thus, we have shown that a construction similar to GLVV \cite{10.1145/2220357.2220363} and \cite{DBLP:journals/corr/GogaczT15} applies to size-bound estimations using the Rényi entropy.

\begin{example}\label{ex:example3}
This example continues \autoref{ex:example1} and \autoref{ex:example2}. Instead of assigning classical probabilities to the tuples in the resulting table, we convert the tables in \autoref{ex:example2} into density matrices. The following matrices demonstrate this idea. In an implementation each variable is encoded into a qubit register, so each matrix would have dimension $2^k \times 2^k$ for a suitable $k$. The elements, such as $|m,1,u\rangle$, represent the fixed basis in \autoref{def:tuples_in_density_matrices}.
 \[
\rho_{xyz} =
\begin{array}{c|ccc}
 & |m,1,u\rangle & |n,2,v\rangle & |m,1,v\rangle \\
\hline
\langle m,1,u| & \frac{1}{3} & 0 & 0 \\
\langle n,2,v| & 0 & \frac{1}{3} & 0 \\
\langle m,1,v| & 0 & 0 & \frac{1}{3} \\
\end{array}
\]

\begin{equation*}
\rho_{xy} =
\begin{array}{c|cc}
 & |m,1\rangle & |n,2\rangle \\
\hline
\langle m,1| & \frac{2}{3} & 0 \\
\langle n,2| & 0 & \frac{1}{3} \\
\end{array}    
\end{equation*}
The previous density matrices are all classical because all the off-diagonal elements are zero. Due to space reasons, we omit $\rho_{xz}$ and $\rho_{yz}$, but they follow the same pattern. If the rows and columns are indexed with the computational basis, then in terms of the pinching map (\autoref{def:pinching_map}), this means that $\Delta(\rho_{i}) = \rho_i$ for each $i \in \left\{ xyz, xy, yz, xz \right\}$.
\end{example}

\subsection{Functional dependencies and classical states}

\begin{definition}[Functional dependency via quantum states]\label{def:functional_dependency_for_states}
A state $\rho$ on $\mathcal{H}$ satisfies a functional dependency $A \to B$ if the reduced states $\rho_{A \cup B}$ and $\rho_{A}$ have the same nonzero eigenvalues, counted with multiplicity.
\end{definition}

\begin{lemma}[Functional dependencies and R\'enyi entropies of classical states]\label{lemma:functional_dependency_with_states}
Let $\alpha \in (0,1) \cup (1,\infty)$ and let $\rho = \sum_t p(t)\,|t\rangle\langle t|$ be a \emph{classical} state with distribution $p$ over the tuple basis, and let $X$ be the random variable distributed by $p$. For variable sets $A, B \subseteq \vr(Q)$ the following are equivalent:
\begin{enumerate}
\item $X$ satisfies the functional dependency $A \to B$ in the sense of Definition~\ref{def:functional_dependency_on_random_variables}: for all $t_1, t_2 \in \supp(p)$, $t_1[A] = t_2[A]$ implies $t_1[B] = t_2[B]$.
\item $H_\alpha(\rho_{A \cup B}) = H_\alpha(\rho_A)$.
\end{enumerate}
Moreover, every classical state satisfies the monotonicity $H_\alpha(\rho_A) \le H_\alpha(\rho_{A \cup B})$.
\end{lemma}

\begin{proof}
We write $p(a,b)$ for the joint distribution on the values of $A \cup B$ and $p_A(a) = \sum_b p(a,b)$ for the marginal distribution. Since $\rho$ is diagonal in the tuple basis by the definition, so are its reduced states, and
\begin{displaymath}
\Tr \rho_A^\alpha = \sum_a \Big(\sum_b p(a,b)\Big)^{\!\alpha},
\qquad
\Tr \rho_{A\cup B}^\alpha = \sum_{a,b} p(a,b)^\alpha.  
\end{displaymath}
Let $\alpha \in (0,1)$. The function $t \mapsto t^\alpha$ is strictly subadditive on $[0,\infty)$: for $s > 0$ the function $g(t) = s^\alpha + t^\alpha - (s+t)^\alpha$ satisfies $g(0) = 0$ and $g'(t) = \alpha\big(t^{\alpha-1} - (s+t)^{\alpha-1}\big) > 0$ for $t > 0$, because $x \mapsto x^{\alpha-1}$ is strictly decreasing. Iterating with subadditivity,
\begin{displaymath}
\Big(\sum_b p(a,b)\Big)^{\!\alpha} \le \sum_b p(a,b)^\alpha    
\end{displaymath}
for every $a$, with equality if and only if at most one $b$ has $p(a,b) > 0$. Summing over $a$ gives $\Tr \rho_A^\alpha \le \Tr \rho_{A\cup B}^\alpha$, with equality if and only if every $a$ in the support of $p_A$ extends to exactly one $b$. This is precisely condition (1). Since $x \mapsto \frac{1}{1-\alpha}\log x$ is strictly increasing for $\alpha < 1$, both the equivalence and the monotonicity follow.

For $\alpha > 1$ the map $t \mapsto t^\alpha$ is strictly superadditive, so all trace inequalities are reversed, and the prefactor $\frac{1}{1-\alpha} < 0$ reverses them back. Thus, the conclusions are identical.
\end{proof}

The following example shows that the assumption that the state is classical is necessary.

\begin{example}[Lemma~\ref{lemma:functional_dependency_with_states} does not extend beyond classical states]
\label{ex:bell}
On two qubits let $\rho = \tfrac12 |\Phi^+\rangle\langle\Phi^+| + \tfrac12 |\Psi^+\rangle\langle\Psi^+|$ with $|\Phi^+\rangle = (|00\rangle + |11\rangle)/\sqrt2$ and $|\Psi^+\rangle = (|01\rangle + |10\rangle)/\sqrt2$. Its spectrum is $\{\tfrac12,\tfrac12,0,0\}$ and both marginals equal $\mathds{I}/2$, so $H_\alpha(\rho_{AB}) = H_\alpha(\rho_A) = \log 2$, for every $\alpha$ i.e. $H_\alpha(\rho_{A \cup B}) = H_\alpha(\rho_A)$. Yet $\Delta(\rho) = \mathds{I}/4$, which indicates that the pinched data has full support $\{00,01,10,11\}$ and satisfies no functional dependency. Hence (2)~$\Rightarrow$~(1) of \autoref{lemma:functional_dependency_with_states} provably fails outside the classical states.
\end{example}

The next result further clarifies the connection between quantum states, classical distributions and functional dependencies.

\begin{theorem}\label{lemma:classical_quantum_functional_dependencies}
Let $X$ be a classical random variable satisfying every functional dependency in $\mathcal{F}$, and let $\rho_X := \sum_t p_X(t)\,|t\rangle\langle t|$ be its encoding (Definition~\ref{def:tuples_in_density_matrices}). Then $\rho_X$ is classical, satisfies every dependency in $\mathcal{F}$ in the sense of \autoref{def:functional_dependency_for_states}, and $H_\alpha\big((\rho_X)_S\big) = H_\alpha(X_S)$ for every $\emptyset \ne S \subseteq \vr(Q)$ and every $\alpha$, where the right-hand side is the classical R\'enyi entropy of the marginal $X_S$.
\end{theorem}
\begin{proof}
By definition of reduced states, the reduced states of $\rho_X$ are the encodings of the marginals of $X$, so all entropies coincide order by order. The dependencies are transferred by Lemma~\ref{lemma:functional_dependency_with_states}, and for classical states, Definition~\ref{def:functional_dependency_for_states} coincides with Definition~\ref{def:functional_dependency_on_random_variables} applied to the support.

\end{proof}



\subsection{Rényi entropy program}

In this subsection, we start deriving the first main result of this paper which is the optimization problem in terms of Rényi entropies. Throughout this subsection, $\Delta_A$, $\Delta_B$ and $\Delta_{AB}$ denote the pinching maps of \autoref{def:pinching_map} with respect to the bases $\{|a\rangle\}$ of $\mathcal{H}_A$, $\{|b\rangle\}$ of $\mathcal{H}_B$, and the product basis $\{|a\rangle \otimes |b\rangle\} =: \{|a,b\rangle\}$ of $\mathcal{H}_A \otimes \mathcal{H}_B$, respectively.

\begin{lemma}[Pinching commutes with the partial trace]\label{lem:pinch}
For every operator $\rho$ on $\mathcal{H}_A \otimes \mathcal{H}_B$ with the product computational basis, we have $\Tr_B \circ\, \Delta_{AB} = \Delta_A \circ \Tr_B $.
\end{lemma}

\begin{proof}
For the left-hand side, $\Delta_{AB}(\rho) = \sum_{a,b} \langle a,b|\rho|a,b\rangle |a,b\rangle\langle a,b|$ is diagonal in the product basis, and $\Tr_B(|a,b\rangle\langle a,b|) = |a\rangle\langle a|$, hence
\begin{displaymath}
    \Tr_B\big(\Delta_{AB}(\rho)\big) = \sum_{a} \Big( \sum_{b} \langle a, b | \rho | a, b \rangle \Big) |a\rangle\langle a| .
\end{displaymath}
For the right-hand side, $\Delta_A$ is diagonal by definition, and by \autoref{lem:ptrace_entries} its surviving entries are $\langle a | \Tr_B(\rho) | a \rangle = \sum_b \langle a,b|\rho|a,b\rangle$, so
\begin{align*}
    \Delta_A\big(\Tr_B(\rho)\big) &= \sum_{a} \langle a | \Tr_B(\rho) | a \rangle\, |a\rangle\langle a| \\ 
    & = \sum_{a} \Big( \sum_{b} \langle a, b | \rho | a, b \rangle \Big)|a\rangle\langle a|.
\end{align*}
The two expressions coincide.
\end{proof}

\begin{theorem}
\label{thm:certificate}
Let $\alpha \in (0,\infty)$ and let $\rho$ be any state on $H$. Then
\begin{displaymath}
    H_\alpha\big(\Delta(\rho)\big) \ge H_\alpha(\rho),
\end{displaymath}
with equality if and only if $\Delta(\rho) = \rho$.
\end{theorem}

\begin{proof}
Let $d$ be the vector of diagonal entries of $\rho$ in the computational basis (equivalently, the spectrum of $\Delta(\rho)$) and $\lambda$ the spectrum of $\rho$. By Schur's theorem (e.g., Theorem 9.B.1~\cite{Marshall_Olkin_Arnold_2011}), we have $d \prec \lambda$ . The R\'enyi entropy is strictly Schur-concave on the probability simplex for every $\alpha \in (0,\infty)$: for $\alpha \ne 1$ the power sum $P_\alpha(x) = \sum_i x_i^\alpha$ is strictly Schur-convex for $\alpha > 1$ and strictly Schur-concave for $\alpha < 1$, because $t \mapsto t^\alpha$ is strictly convex resp.\ strictly concave (e.g. 3.C.1~\cite{Marshall_Olkin_Arnold_2011}), and $H_\alpha = \frac{1}{1-\alpha}\log_b P_\alpha$ composes to a strictly Schur-concave function for all $\alpha$:
\begin{itemize}
    \item For $\alpha > 1$: $P_\alpha(x)$ is strictly Schur-convex. Because the logarithm is strictly increasing (assuming base $b > 1$), $\log_b P_\alpha$ remains strictly Schur-convex. Multiplying by $\frac{1}{1-\alpha}$ (which is negative) flips the inequality, making the entire expression strictly Schur-concave.
    \item For $0 < \alpha < 1$: $P_\alpha(x)$ is strictly Schur-concave. The logarithm preserves this concavity, and multiplying by $\frac{1}{1-\alpha}$ (which is positive) maintains it, keeping the expression strictly Schur-concave.
    \item For $\alpha = 1$: The limit yields the Shannon entropy, which is standard (``classical'') and known to be strictly Schur-concave.
\end{itemize}
Hence $H_\alpha(\Delta\rho) = H_\alpha(d) \ge H_\alpha(\lambda) = H_\alpha(\rho)$.

Suppose equality holds. Strict Schur-concavity forces $d$ to be a permutation of $\lambda$, so $\sum_i d_i^2 = \sum_i \lambda_i^2 = \Tr \rho^2$. But
\begin{displaymath}
    \Tr \rho^2 = \sum_{i,j} |\rho_{ij}|^2 = \sum_i d_i^2 + \sum_{i \ne j} |\rho_{ij}|^2,
\end{displaymath}
so all off-diagonal entries vanish and $\Delta(\rho) = \rho$. The converse of the theorem is obvious.
\end{proof}

Next, we present the corresponding theorem to \autoref{thm:glvv}, except that the result is expressed in terms of Rényi entropies and quantum states. This is now the first main result in this work.

\begin{theorem}
\label{thm:main_result}
For a state $\rho$ on $\mathcal{H}$ the following are equivalent:
\begin{enumerate}
    \item $\Delta(\rho_S) = \rho_S$ for all $\emptyset \ne S \subseteq \vr(Q)$
    \item $\Delta(\rho) = \rho$
    \item $H_\beta(\Delta(\rho)) = H_\beta(\rho)$ for some, equivalently for every, $\beta \in (0,\infty)$.
\end{enumerate}

Now, restricting to $\alpha \in (0,1)$, we obtain the following program:
\begin{displaymath}
\begin{array}{lll}
\text{maximize}   & H_\alpha(\rho_{u_0}) & \\[2pt]
\text{subject to} & H_\alpha(\rho_{u_j}) \le 1 & \forall j \in [m] \\[2pt]
                  & H_\alpha(\rho_{A \cup B}) = H_\alpha(\rho_A) & \text{for each f.d. } A \to B \in \mathcal{F} \\[2pt]
                  & H_\alpha(\Delta(\rho)) = H_\alpha(\rho) , &
\end{array}    
\end{displaymath}
in which every constraint is an equality or inequality between Rényi entropies. We write $h_\alpha(Q)$ for the optimal value of the previous program. By \autoref{thm:certificate}, the last constraint restricts the optimization to classical states, and on classical states the equality constraints hold if and only if the support of the encoded distribution satisfies the dependencies in $\mathcal{F}$, by \autoref{lemma:functional_dependency_with_states}.
\end{theorem}

\begin{proof}
(1)$\Rightarrow$(2): We take $S = \vr(Q)$. (2)$\Rightarrow$(1): We iterate Lemma~\ref{lem:pinch}, which yields $\rho_S = \Tr_{\bar S}\,\rho = \Tr_{\bar S}\,\Delta\rho = \Delta(\Tr_{\bar S}\,\rho) = \Delta(\rho_S)$. (2)$\Leftrightarrow$(3): This follows immediately from Theorem~\ref{thm:certificate}.
\end{proof}

%
%

\autoref{def:renyi_entropy} leaves the base $b > 1$ of the logarithm unspecified. To relate the program of \autoref{thm:main_result} to a concrete database instance $D$, we choose $b := \rmx(D)$. This means that we compute every R\'enyi entropy as $H_\alpha(\rho) = \frac{1}{1-\alpha} \log_{\rmx(D)} \Tr\rho^{\alpha}$. This is the same normalization that underlies the classical program of \autoref{thm:glvv}. There, the constraint $h(u_j) \le 1$ expresses that every entropy has implicitly been divided by $\log \rmx(D)$. With this convention, the program of \autoref{thm:main_result} is sound, meaning that every database instance realizes its own worst-case exponent as the objective value of some feasible state.

\begin{corollary}[Soundness of the R\'enyi program]\label{cor:soundness}
Fix $\alpha \in (0,1)$, let $Q = \mathrm{chase}(Q) = R_0(u_0) \leftarrow R_{i_1}(u_1) \wedge \ldots \wedge R_{i_m}(u_m)$ be a conjunctive query with functional dependencies $\mathcal{F}$, and let $h_\alpha(Q)$ be the optimal value of the program in \autoref{thm:main_result}, with entropies computed in base $\rmx(D)$ for the database $D$.
\begin{enumerate}
    \item Let $D$ be a database satisfying $\mathcal{F}$ with $\rmx(D) \ge 2$ and $Q(D) \ne \emptyset$. Then there exists a state $\rho$ that is feasible for the program and attains the objective value
    \begin{displaymath}
        H_\alpha(\rho_{u_0}) = \log_{\rmx(D)} |Q(D)| = s(Q, D).
    \end{displaymath}
    \item Consequently, $s(Q, D) \le h_\alpha(Q)$ for every database $D$ as in part (1), and
    \begin{displaymath}
        |Q(D)| \le \rmx(D)^{\,h_\alpha(Q)}
    \end{displaymath}
    for every database $D$ satisfying $\mathcal{F}$.
\end{enumerate}
\end{corollary}

\begin{proof}
If $Q(D) = \emptyset$ or $\rmx(D) \le 1$, then $|Q(D)| \le 1$ and the bound is trivial. So assume $\rmx(D) \ge 2$ and $Q(D) \ne \emptyset$, and write $b := \rmx(D)$.
 
For each $y \in Q(D)$ we fix a so-called witness $t_y \in Q'(D)$ with $t_y[u_0] = y$. Let $X$ be uniformly distributed on $T := \{ t_y \colon y \in Q(D) \}$. Because $t_y[u_0] = y$, the projection $t \mapsto t[u_0]$ maps $T$ bijectively onto $Q(D)$. Hence $|T| = |Q(D)|$ and the marginal $X_{u_0}$ is uniform on $Q(D)$. Moreover, $X$ satisfies every dependency $A \to B \in \mathcal{F}$: the dependency is given by a body atom $R_{i_j}(u_j)$ (\autoref{sec:prerequisites}), and two tuples of $Q'(D) \supseteq T$ that agree on $A$ restrict at this atom to tuples of $R_{i_j}(D)$ that agree on the $A$-positions and therefore, since $D$ satisfies $\mathcal{F}$, also on the $B$-positions.
 
Let $\rho$ be the encoding of $X$ from \autoref{def:tuples_in_density_matrices}. By \autoref{lemma:classical_quantum_functional_dependencies}, $\rho$ is classical and $H_\alpha(\rho_S) = H_\alpha(X_S)$ for every $\emptyset \ne S \subseteq \vr(Q)$. Feasibility can be shown in the following way. The constraint $H_\alpha(\Delta(\rho)) = H_\alpha(\rho)$ holds by \autoref{thm:main_result}, the dependency constraints hold by \autoref{lemma:functional_dependency_with_states}, and for every $j \in [m]$ the marginal $X_{u_j}$ is supported on at most $|R_{i_j}(D)| \le b$ points (reading values along the list $u_j$ embeds $\pi_{u_j}(T)$ into $R_{i_j}(D)$) so the computations at the beginning of this section give $H_\alpha(\rho_{u_j}) \le \log_b |R_{i_j}(D)| \le 1$. By the same computations, since $X_{u_0}$ is uniform on $|Q(D)|$ points, $H_\alpha(\rho_{u_0}) = \log_b |Q(D)| = s(Q, D)$. As $h_\alpha(Q)$ is the supremum of the objective over the feasible states, $s(Q, D) \le h_\alpha(Q)$, and exponentiating in base $b$ yields $|Q(D)| \le \rmx(D)^{h_\alpha(Q)}$.
\end{proof}

\subsection{Dichotomy of Rényi entropy program}

In this subsection, we focus on the second main result of this work, which clarifies the limits of the optimization problem in \autoref{thm:main_result}. The limits can be understood in light of a dichotomy result: either we obtain a result that aligns with the coloring number, as defined in GLVV~\cite{10.1145/2220357.2220363}, or the bound increases without bound. The following definitions are from GLVV~\cite{10.1145/2220357.2220363}.

\begin{definition}[Coloring]
\label{def:coloring}
Let $Q = R_0(u_0) \leftarrow R_{i_1}(u_1) \wedge \cdots \wedge R_{i_m}(u_m)$ be a conjunctive query with functional dependencies $\mathcal{F}$. A \emph{valid coloring} of $Q$ with color set $[d]$ is a mapping $\mathcal{L} \colon \vr(Q) \to 2^{[d]}$ assigning to each variable a (possibly empty) set of colors, such that
\begin{enumerate}
    \item for every functional dependency $A \to B \in \mathcal{F}$,  we have $\mathcal{L}(B) \subseteq \bigcup_{X \in A} \mathcal{L}(X)$, and
    \item $\mathcal{L}(X) \neq \emptyset$ for at least one $X \in \vr(Q)$.
\end{enumerate}
\end{definition}

\begin{definition}[Color number]
\label{def:color_number}
The \emph{color number} of a valid coloring $\mathcal{L}$ of $Q$ is the ratio of the number of colors occurring in the head to the largest number of colors occurring in a single body atom. The \emph{color number of $Q$} is
\begin{displaymath}
    C(Q) :=\max_{\mathcal{L}} \;
    \frac{\big| \cup_{X \in u_0} \mathcal{L}(X) \big|}
         {\max_{j \in [m]} \big| \cup_{X \in u_j} \mathcal{L}(X) \big|},
\end{displaymath}
where $\mathcal{L}$ ranges over all valid colorings of $Q$.
\end{definition}

\begin{definition}[Closure under functional dependencies]\label{def:closure}
For $V \subseteq \vr(Q)$, the closure $\clF(V)$ is the smallest subset of variables $W \supseteq V$ such that $A \to B \in \mathcal{F}$ and $A \subseteq W$ imply $B \in W$. We say that $V$ is closed under $\mathcal{F}$ if $V = \clF(V)$.
\end{definition}

\begin{definition}[Head absorption]\label{def:absorption}
Query $Q$ satisfies \emph{head absorption} if $\vr(u_0) \subseteq \clF(\vr(u_j))$ for some $j \in [m]$.
\end{definition}

\begin{example}\label{ex:example4}
Next, we continue our running example. For the triangle query $Q(x,y,z) \leftarrow R_1(x,y) \wedge R_2(y,z) \wedge R_3(x,z)$ with $\mathcal{F} = \emptyset$, the coloring $\mathcal{L}(x) = \{1\}$, $\mathcal{L}(y) = \{2\}$, $\mathcal{L}(z) = \{3\}$ is valid. This follows from the fact that the head carries three colors and every body atom carries two, so its color number is $3/2$ and $C(Q) = 3/2$, which matches the classical bound computed in \autoref{ex:example1}. The head absorption property fails, which is in agreement with \autoref{lem:coloring}. This example is also in alignment with the other main results, dichotomy \autoref{thm:dichotomy}.
\end{example}

The following lemma is used multiple times in the following proofs.

\begin{lemma}[Induction on closure sets]\label{lem:closure_induction}
Let $V \subseteq \vr(Q)$ and define the sets $W_0 := V$ and $W_{k+1} := W_k \cup \{ B \colon A \to B \in \mathcal{F},\, A \subseteq W_k \}$ for $k \ge 0$. Then $\clF(V) = \bigcup_{k \ge 0} W_k$. If a set $W \subseteq \vr(Q)$ contains $V$ and $W$ is closed under $\mathcal{F}$, then $\clF(V) \subseteq W$.
\end{lemma}

\begin{proof}
Since $\vr(Q)$ is finite and the sets $W_k$ are increasing, the sequence stabilizes: there is $K \le |\vr(Q)|$ with $W_{K+1} = W_K$, and hence $\bigcup_{k \ge 0} W_k = W_K$. We first check that $W_K$ contains $V$ and is closed under $\mathcal{F}$. It contains $V = W_0$ by construction, and if $A \to B \in \mathcal{F}$ with $A \subseteq W_K$, then $B \in W_{K+1} = W_K$. Next, we show that $W_K$ is contained in every set $W$ that contains $V$ and is closed under $\mathcal{F}$. We prove $W_k \subseteq W$ by induction on $k$. The base case $W_0 = V \subseteq W$ holds by assumption. For the induction step, assume $W_k \subseteq W$ and let $B \in W_{k+1}$ be added on account of a functional dependency $A \to B \in \mathcal{F}$ with $A \subseteq W_k$. Then $A \subseteq W$, and the closedness of $W$ gives $B \in W$. Hence $W_K \subseteq W$. Thus, $W_K$ is the smallest set that contains $V$ and is closed under $\mathcal{F}$, which is precisely the definition of $\clF(V)$ in \autoref{def:closure}. Both claims follow.
\end{proof}

The following lemma connects the color number and head absorption.

\begin{lemma}[Closure coloring]\label{lem:coloring}
Head absorption fails if and only if $C(Q) > 1$. In that case, $C(Q) \ge m/(m-1)$.
\end{lemma}

\begin{proof}
Assume first that head absorption holds for an index $j \in [m]$, and let $\mathcal{L}$ be any valid coloring. Consider the set
\begin{displaymath}
    W := \Big\{ X \in \vr(Q) \colon \mathcal{L}(X) \subseteq
    \cup_{Y \in \vr(u_j)} \mathcal{L}(Y) \Big\}
\end{displaymath}
of variables whose colors already occur on the atom $u_j$. The set $W$ contains $\vr(u_j)$, since for $X \in \vr(u_j)$ the set $\mathcal{L}(X)$ is one of the sets in the union. The set $W$ is also closed under $\mathcal{F}$: if $A \to B \in \mathcal{F}$ and $A \subseteq W$, then the coloring condition of \autoref{def:coloring} gives $\mathcal{L}(B) \subseteq \bigcup_{X \in A} \mathcal{L}(X) \subseteq \bigcup_{Y \in \vr(u_j)} \mathcal{L}(Y)$, so $B \in W$. By \autoref{lem:closure_induction}, $\clF(\vr(u_j)) \subseteq W$. Since head absorption gives $\vr(u_0) \subseteq \clF(\vr(u_j)) \subseteq W$, taking the union over the head variables yields $\bigcup_{X \in u_0} \mathcal{L}(X) \subseteq \bigcup_{Y \in u_j} \mathcal{L}(Y)$. Hence the numerator of the color number of $\mathcal{L}$ is at most $\big| \bigcup_{X \in u_j} \mathcal{L}(X) \big|$, which is at most the denominator, so the color number of $\mathcal{L}$ is at most $1$. As $\mathcal{L}$ was arbitrary, $C(Q) \le 1$.

Conversely, assume that head absorption fails. Then we actually have $m \ge 2$: if $m = 1$, then every head variable occurs in the single body atom $u_1$, so
$\vr(u_0) \subseteq \vr(u_1) \subseteq \clF(\vr(u_1))$ and absorption would hold. We define a coloring with color set $[m]$ by
\begin{displaymath}
    \mathcal{L}(X) := \{ j \in [m] \colon X \notin \clF(\vr(u_j))\},
\end{displaymath}
so that the color $j$ is given by exactly those variables that the atom $u_j$ does not determine.

We first argue that $\mathcal{L}$ is a valid coloring. Let $A \to B \in \mathcal{F}$ and let $j \in \mathcal{L}(B)$, that is, $B \notin
\clF(\vr(u_j))$. If every $X \in A$ were to lie in $\clF(\vr(u_j))$, then $A \subseteq \clF(\vr(u_j))$, and since $\clF(\vr(u_j))$ is closed under
$\mathcal{F}$ we would get $B \in \clF(\vr(u_j))$, which is a contradiction. Hence some $X \in A$ satisfies $X \notin \clF(\vr(u_j))$, i.e.\ $j \in \mathcal{L}(X)$,
which establishes $\mathcal{L}(B) \subseteq \bigcup_{X \in A}\mathcal{L}(X)$. Moreover, because absorption fails at the index $j$, there exists a head
variable $X_j \in \vr(u_0)$ with $X_j \notin \clF(\vr(u_j))$, so $j \in \mathcal{L}(X_j)$ and $\mathcal{L}$ is not identically empty.

It remains to bound the color number of $\mathcal{L}$ from below. For the numerator, the variables $X_1, \ldots, X_m$ of the previous paragraph all lie
in $\vr(u_0)$ and witness the colors $1, \ldots, m$ respectively, so the head carries all $m$ colors and $\big| \bigcup_{X \in u_0}\mathcal{L}(X) \big| = m$.
For the denominator, fix $i \in [m]$. Then, every $X \in \vr(u_i)$ lies in $\vr(u_i) \subseteq \clF(\vr(u_i))$, so $i \notin \mathcal{L}(X)$, and
therefore the color $i$ appears nowhere on $u_i$, giving $\big| \bigcup_{X \in u_i}\mathcal{L}(X) \big| \le m - 1$. The denominator is
also at least $1$ which can be seen as follows. The head variable $X_1$ occurs in some body atom, since every head variable occurs in the body, and it carries the color $1$. Hence the color number of $\mathcal{L}$ is at least $\tfrac{m}{m-1}$, and $C(Q) \ge \tfrac{m}{m-1} > 1$ because $m \ge 2$.

The two halves together prove the equivalence: absorption implies $C(Q) \le 1$, and its failure implies $C(Q) \ge \tfrac{m}{m-1} > 1$.
\end{proof}

Now we are ready to prove the second main result in this paper which clarifies the limits of Rényi entropy.

\begin{theorem}[Dichotomy result]\label{thm:dichotomy}
Fix $\alpha \in (0,1)$. Let $Q = \mathrm{chase}(Q)$ be a query with functional dependencies $\mathcal{F}$, and let $h_\alpha(Q)$ be the optimal value of the program in \autoref{thm:main_result}.
    \begin{enumerate}
        \item If $Q$ satisfies head absorption, then $h_\alpha(Q) \le 1$. Moreover $|Q(D)| \le \rmx(D)$ for every database $D$ satisfying $\mathcal{F}$.
        \item Otherwise $h_\alpha(Q) = +\infty$.
    \end{enumerate}
\end{theorem}

\begin{proof}
\emph{(1)} Let $\rho$ be any feasible state of the program in \autoref{thm:main_result}. By \autoref{thm:certificate}, the last constraint forces $\Delta(\rho) = \rho$, so $\rho = \sum_t p(t)|t\rangle\langle t|$ is classical with a distribution $p$ over the tuple basis. Since $\rho$ is classical and satisfies the entropic functional dependency constraints, \autoref{lemma:functional_dependency_with_states}, applied in the direction (2) $\Rightarrow$ (1) to each $A \to B \in \mathcal{F}$, shows that $\supp(p)$ satisfies every functional dependency in $\mathcal{F}$.

Let $j \in [m]$ be an index witnessing head absorption, so that $\vr(u_0) \subseteq \clF(\vr(u_j))$. We claim that $\supp(p)$ also satisfies the derived dependency $\vr(u_j) \to \vr(u_0)$, which need not belong to $\mathcal{F}$. Let $t_1, t_2 \in \supp(p)$ with $t_1[u_j] = t_2[u_j]$, and consider their agreement set
\begin{displaymath}
    W := \{ X \in \vr(Q) \colon t_1[X] = t_2[X] \}.
\end{displaymath}
The set $W$ contains $\vr(u_j)$ because $t_1[u_j] = t_2[u_j]$. The set $W$ is also closed under $\mathcal{F}$: if $A \to B \in \mathcal{F}$ and $A \subseteq W$, then $t_1[A] = t_2[A]$, and since $\supp(p)$ satisfies the dependency $A \to B$, we obtain $t_1[B] = t_2[B]$, that is, $B \subseteq W$. By \autoref{lem:closure_induction}, $\clF(\vr(u_j)) \subseteq W$, and head absorption gives $\vr(u_0) \subseteq W$; that is, $t_1[u_0] = t_2[u_0]$, which proves the claim.

The claim is precisely condition (1) of \autoref{lemma:functional_dependency_with_states} for the variable sets $A = \vr(u_j)$ and $B = \vr(u_0)$. The lemma therefore yields $H_\alpha(\rho_{u_j \cup u_0}) = H_\alpha(\rho_{u_j})$, and its monotonicity statement gives $H_\alpha(\rho_{u_0}) \le H_\alpha(\rho_{u_0 \cup u_j})$. Combining these with the feasibility constraint carried by the atom $u_j$,
\begin{displaymath}
    H_\alpha(\rho_{u_0})
    \le H_\alpha(\rho_{u_j \cup u_0})
    = H_\alpha(\rho_{u_j}) \le 1 .
\end{displaymath}
As $\rho$ was an arbitrary feasible state and $h_\alpha(Q)$ is the supremum of the objective over the feasible states, $h_\alpha(Q) \le 1$. The database statement follows directly from \autoref{cor:soundness}: $|Q(D)| \le \rmx(D)^{h_\alpha(Q)} \le \rmx(D)$.

\emph{(2)} Assume head absorption fails. By \autoref{lem:coloring}, $C(Q) > 1$, so by \autoref{def:color_number} there exists a valid coloring whose color number exceeds $1$. Fix such a coloring $\mathcal{L}$ and discard from its color set every color that no variable carries; this changes neither the numerator nor the denominator of the color number. Let $[d]$ denote the remaining color set, and define
\begin{displaymath}
    a := \Big|\bigcup_{X \in u_0} \mathcal{L}(X)\Big|,
    \qquad
    r := \max_{j \in [m]} \Big|\bigcup_{X \in u_j} \mathcal{L}(X)\Big|.
\end{displaymath} 
First, we note that $r \ge 1$. By condition (2) of \autoref{def:coloring} at least one color is carried, every variable of $Q$ occurs in some body atom, and hence every carried color occurs on some body atom. Since the color number of $\mathcal{L}$ is $a/r > 1$ and both quantities are integers, $a \ge r + 1$.

The proof proceeds in three steps. First, we construct, for every integer $M \ge 2$, a set $T_M$ of tuples that satisfies all functional dependencies and whose head projection ($M^a$ tuples) is strictly larger than any body projection (at most $M^r$ tuples). Second, we place a carefully chosen distribution $p^{(M)}$ on $T_M$. Third, we verify that the corresponding classical state is feasible for the program for every $M$, while its objective value grows without bound as $M \to \infty$. Since $h_\alpha(Q)$ is the supremum of the objective over the feasible states, this proves $h_\alpha(Q) = +\infty$.

\emph{Step 1: constructing product table $T_M$.} Fix an integer $M \ge 2$ and write $[M] = \{1, \ldots, M\}$. We consider each color $i \in [d]$ as an independent degree of freedom taking values in $[M]$, and of a variable $X$ as recording the values of exactly those colors it carries. Formally, for $c = (c_1, \ldots, c_d) \in [M]^d$ define the tuple $t_c$ over $\vr(Q)$ by $t_c[X] := (c_i)_{i \in \mathcal{L}(X)}$ where the coordinates are listed in increasing order of $i$, and where $t_c[X]$ is a fixed null symbol when $\mathcal{L}(X) = \emptyset$. Define $T_M := \{ t_c \colon c \in [M]^d \}$.

The table $T_M$ has the properties promised above. First, since every color in $[d]$ is carried by some variable, the map $c \mapsto t_c$ is injective, so $|T_M| = M^d$. Second, $T_M$ satisfies every functional dependency $A \to B \in \mathcal{F}$: the restriction $t_c[A]$ determines the coordinates $c_i$ for every $i \in \bigcup_{X \in A}\mathcal{L}(X)$, and validity of the coloring gives $\mathcal{L}(B) \subseteq \bigcup_{X \in A}\mathcal{L}(X)$, so $t_c[A]$ determines $t_c[B]$. Third, $t_c[u_0]$ consists exactly of the coordinates $c_i$ with $i \in \bigcup_{X \in u_0}\mathcal{L}(X)$, so $|\pi_{u_0}(T_M)| = M^{a}$ and every point of $\pi_{u_0}(T_M)$ has exactly $M^{d-a}$ preimages in $T_M$. Similarly $t_c[u_j]$ is determined by the coordinates with $i \in \bigcup_{X \in u_j}\mathcal{L}(X)$, so $|\pi_{u_j}(T_M)| \le M^{r}$ for every $j \in [m]$.

\emph{Step 2: distribution on $T_M$.} We now choose the distribution. The choice is guided by the following observation about the Rényi entropy for $\alpha \in (0,1)$. Suppose a total probability mass $\varepsilon$ is spread uniformly over $N$ points, each point receiving mass $\varepsilon/N$. The contribution of these points to the power sum $\Tr \rho^\alpha$ is
\begin{displaymath}
    N \cdot \Big( \frac{\varepsilon}{N} \Big)^{\!\alpha}
    = \varepsilon^{\alpha} N^{1-\alpha}.
\end{displaymath}
Because $\alpha < 1$, this quantity grows with $N$ even when $\varepsilon$ is small: the power sum rewards the size of the support much more than the mass placed on it. We use this asymmetry as follows. We place almost all mass on one fixed tuple, which contributes essentially nothing to any power sum, and spread the small remaining mass $\varepsilon$ uniformly over all of $T_M$. On a body atom $u_j$, this remaining mass lands on at most $M^r$ points, and we will choose $\varepsilon$ so small that the resulting contribution $\varepsilon^\alpha M^{r(1-\alpha)}$ stays constant in $M$. On the head $u_0$, however, the same mass lands on $M^a$ points, and since $a > r$, the head contribution $\varepsilon^\alpha M^{a(1-\alpha)}$ then diverges as $M \to \infty$. This is where the strict inequality of \autoref{lem:coloring} enters.

We make this precise. Let $b > 1$ be the base of the logarithm in \autoref{def:renyi_entropy}. Because $\frac{1}{1-\alpha} > 0$ for $\alpha \in (0,1)$ and $x \mapsto \frac{1}{1-\alpha}\log_b x$ is strictly increasing, the feasibility constraint $H_\alpha(\rho_{u_j}) \le 1$ is equivalent to the power sum constraint
\begin{equation}\label{eq:feasibility_powersum}
    \Tr \big( \rho_{u_j} \big)^\alpha \;\le\; b^{1-\alpha}.
\end{equation}
Since $b^{1-\alpha} > 1$ and a point mass has power sum equal to $1$, the constraint \eqref{eq:feasibility_powersum} leaves a positive budget of $b^{1-\alpha} - 1$ above the point mass. We reserve half of this budget for the uniformly spread mass by defining the constant
\begin{displaymath}
    c_0 := \Big( \tfrac{1}{2} \big( b^{1-\alpha} - 1 \big) \Big)^{1/\alpha} > 0 .
\end{displaymath}
Now fix a tuple $t_0 \in T_M$, write $\delta_{t_0}$ for the point mass at $t_0$, and let $U$ be the uniform distribution on $T_M$, which assigns mass $M^{-d}$ to each of the $M^d$ tuples of $T_M$. We define
\begin{displaymath}
    p^{(M)} := (1-\varepsilon)\, \delta_{t_0} + \varepsilon\, U
\end{displaymath}
where $\varepsilon := c_0 \, M^{-r(1-\alpha)/\alpha}$. The exponent of $M$ is chosen exactly so that $\varepsilon^{\alpha} M^{r(1-\alpha)} = c_0^{\alpha}$ holds for every $M$ and this is necessarily its only purpose. Since $r \ge 1$ and $\alpha \in (0,1)$, the exponent $-r(1-\alpha)/\alpha$ is negative, so $\varepsilon \to 0$ as $M \to \infty$ and $\varepsilon \in (0,1)$ for all sufficiently large $M$. We restrict to such $M$. Finally, let $$\rho^{(M)} := \sum_{t \in T_M} p^{(M)}(t)\, |t\rangle\langle t|$$ be the classical state encoding $p^{(M)}$ in the sense of \autoref{def:tuples_in_density_matrices}.

\emph{Step 3a: $\rho^{(M)}$ is feasible.} The state $\rho^{(M)}$ is classical by construction, so the constraint $H_{\alpha}(\Delta(\rho)) = H_{\alpha}(\rho)$ holds. Its support is $\supp(p^{(M)}) = T_M$, which satisfies every functional dependency in $\mathcal{F}$ by Step~1, so every functional dependency constraint holds by \autoref{lemma:functional_dependency_with_states}. It remains to verify \eqref{eq:feasibility_powersum} for every body atom $u_j$.

Fix $j \in [m]$. Taking the marginal of $p^{(M)}$ on the variables of $u_j$ gives
\begin{displaymath}
    p^{(M)}_{u_j} = (1-\varepsilon)\, \delta_{t_0[u_j]} + \varepsilon\, q_j ,
\end{displaymath}
where $q_j$ denotes the marginal of $U$ on $u_j$, a distribution supported on $\pi_{u_j}(T_M)$, hence on at most $M^r$ points by Step~1. Note that the point $t_0[u_j]$ lies in the support of $q_j$, so the two terms overlap and the power sum of the mixture is not simply the sum of the power sums. Instead, we bound it using the subadditivity of $t \mapsto t^{\alpha}$, applied entrywise: for every $y$,
\begin{displaymath}
    \big( (1-\varepsilon)\,\delta_{t_0[u_j]}(y) + \varepsilon\, q_j(y) \big)^{\alpha}
    \le (1-\varepsilon)^{\alpha}\, \delta_{t_0[u_j]}(y) + \varepsilon^{\alpha} q_j(y)^{\alpha},
\end{displaymath}
where we also used $\delta_{t_0[u_j]}(y)^\alpha = \delta_{t_0[u_j]}(y) \in \{0,1\}$. Summing over $y$, we obtain
\begin{displaymath}
    \Tr \big(\rho^{(M)}_{u_j}\big)^\alpha
    \le (1-\varepsilon)^{\alpha} + \varepsilon^{\alpha} \sum_{y} q_j(y)^{\alpha}
    \le 1 + \varepsilon^{\alpha} M^{r(1-\alpha)},
\end{displaymath}
where the last step uses $(1-\varepsilon)^\alpha \le 1$ together with the fact that a distribution on at most $N$ points has power sum at most $N^{1-\alpha}$, attained by the uniform distribution. This is also the familiar statement that $H_\alpha$ is maximized by the maximally mixed state, applied with $N = M^r$. By the choice of $\varepsilon$, we have $\varepsilon^{\alpha} M^{r(1-\alpha)} = c_0^{\alpha} = \tfrac12(b^{1-\alpha}-1)$, and therefore
\begin{displaymath}
    \Tr \big(\rho^{(M)}_{u_j}\big)^\alpha
    \le 1 + \tfrac12\big(b^{1-\alpha} - 1\big)
    = \frac{1 + b^{1-\alpha}}{2}
    < b^{1-\alpha} .
\end{displaymath}
This is \eqref{eq:feasibility_powersum}, so $H_\alpha(\rho^{(M)}_{u_j}) < 1$ for every $j$, and $\rho^{(M)}$ is feasible for every sufficiently large $M$.

\emph{Step 3b: the objective diverges.} Next, we bound the objective value $H_\alpha(\rho^{(M)}_{u_0})$ from below. By Step~1, the marginal of $U$ on the head variables is the uniform distribution on the $M^a$ points of $\pi_{u_0}(T_M)$, because all preimages of $\pi_{u_0}$ have the same size. Hence the head marginal of $p^{(M)}$ assigns mass exactly $\varepsilon/M^a$ to each of the $M^a - 1$ points of $\pi_{u_0}(T_M) \setminus \{t_0[u_0]\}$, and mass $(1-\varepsilon) + \varepsilon/M^a$ to the remaining point $t_0[u_0]$. Discarding the contribution of $t_0[u_0]$, we obtain the lower bound
\begin{align*}
    \Tr \big(\rho^{(M)}_{u_0}\big)^\alpha
    &\ge \big(M^a - 1\big) \Big( \frac{\varepsilon}{M^a} \Big)^{\!\alpha}
    \ge \frac{M^a}{2}\cdot \frac{\varepsilon^{\alpha}}{M^{a\alpha}}\\
    &= \tfrac12\, \varepsilon^{\alpha} M^{a(1-\alpha)}
    = \tfrac12\, c_0^{\alpha}\, M^{(a-r)(1-\alpha)},
\end{align*}
where the second inequality uses $M^a - 1 \ge M^a/2$, valid since $M^a \ge 2$, and the last equality substitutes the definition of $\varepsilon$. Applying the increasing map $x \mapsto \frac{1}{1-\alpha}\log_b x$ to both sides gives
\begin{align*}
    H_\alpha\big(\rho^{(M)}_{u_0}\big)
    &\ge \frac{1}{1-\alpha}
      \Big( (a-r)(1-\alpha)\log_b M + \alpha \log_b c_0 - \log_b 2 \Big) \\
    &= (a - r)\log_b M - \kappa_\alpha,
\end{align*}
where $\kappa_\alpha := \frac{\log_b 2 - \alpha \log_b c_0}{1-\alpha},$ which is a finite constant that does not depend on $M$. Since $a - r \ge 1$, the right-hand side tends to $+\infty$ as $M \to \infty$. Thus, the states $\rho^{(M)}$ are feasible for all large $M$ while their objective values are unbounded, and therefore $h_\alpha(Q) = +\infty$.
\end{proof}

Consequently, $h_\alpha(Q) \in [0,1] \cup \{+\infty\}$ for $\alpha \in (0,1)$. This means that the value of the program records whether a size increase is possible, and, unfortunately, it provides no quantitative information about the worst-case exponent $s(Q)$ in Equation~\eqref{eq:sQ}.
\section{Discussion}\label{sec:discussion}

We briefly discuss the obtained results. Reformulating the size-bound problem in terms of Rényi entropy does not reduce the difficulty; rather, it shifts the question from characterizing the Shannon entropy cone to "what are the classical states?" In quantum entropy terms, this problem is centered around characterizing those states for which $H_{\alpha}(\Delta(\rho)) = H_{\alpha}(\rho)$. This is a fundamentally different question from characterizing the Shannon entropy region, and it hopefully brings a new perspective to the problem. Moreover, the dichotomy theorem quantifies the price of this relaxation for fixed $\alpha$. The current Rényi program in \autoref{thm:main_result} is a relaxation of the entropic bound, and thus it cannot be tighter.

This work concludes that the current formulation offers little for classical data. Off-diagonal states $\rho$ can satisfy all entropic functional dependency constraints, but the corresponding pinched states $\Delta(\rho)$ might not. Nevertheless, there are works where databases are used as quantum simulators or optimizers in quantum computational pipelines~\cite{10.1145/3722212.3725126,uotila2025zxdbgraphdatabasequantum,moflic2025ultralargescalecompilationmanipulationquantum,10.1145/3736393.3736694}. We believe that in these cases, one might need to store full density matrices or equivalent representations in a database, and one can view these as representations of quantum data. In these instances, quantum information-theoretic approaches should become increasingly relevant. 
 
\section{Conclusion}

In this work, we presented a quantum information-theoretical formulation to express size bounds for conjunctive queries in relational databases with the presence of arbitrary functional dependencies. The bound is expressed in terms of quantum states and Rényi entropy. The bound induces an optimization formulation whose solution is a sound upper bound for the worst-case size increases. Additionally, we provided a dichotomy theorem that characterizes the bound as at most one or unbounded in the case of quantum states and Rényi entropy of order $\alpha \in (0, 1)$.  The key motivation to use quantum entropies for this problem is the strikingly simple entropy cone for Rényi entropy compared to the classical Shannon entropy. The intersection of quantum information and database theories appears as a promising research direction. In future research, we believe that quantum information theory and degree sequences~\cite{Zhang_Mayer_Khamis_Olteanu_Suciu_2025,Khamis_Nakos_Olteanu_Suciu_2024,khamis2025informationtheorystrikesback} might have an interesting connection to study further.

\balance
\bibliographystyle{ACM-Reference-Format}
\bibliography{sample}

\end{document}